\documentclass{du-journals}
\usepackage{caption} 
\usepackage{rotating}
\journal{Iranian Journal of Astronomy and Astrophysics}
\title{Hierarchical Classification of Variable Stars Using Deep Convolutional Neural Networks}
\author[1]{Mahdi Abdollahi}
\address[1]{School~of~Astronomy,~Institute~for Research~in~Fundamental~Sciences~(IPM),~P. O. Box19395-5531,~Tehran,~Iran; \\ email: m.abdollahi@ipm.ir}
\author[2]{Nooshin Torabi}
\address[2]{Department~of~Physics,~Sharif~University~of~Technology,~P. O. Box 11365-9161,~Tehran,~Iran; \\ email: nooshin\_torabi@physics.sharif.edu}
\author[3]{Sadegh Raeisi}
\address[3]{Department~of~Physics,~Sharif~University~of~Technology,~P. O. Box 11365-9161,~Tehran,~Iran; \\ email: sraeisi@sharif.edu}
\author[4]{Sohrab Rahvar}
\address[4]{Department~of~Physics,~Sharif~University~of~Technology,~P. O. Box11365-9161,~Tehran,~Iran; \\ email: rahvar@sharif.edu}
\begin{document}
\begin{abstract}
The importance of using fast and automatic methods to classify variable stars for large amounts of data is undeniable. There have been many attempts to classify variable stars by traditional algorithms like Random Forest. In recent years, neural networks as classifiers have come to notice because of their lower computational cost compared to traditional algorithms. This paper uses the Hierarchical Classification technique, which contains two main steps of predicting class and then subclass of stars. All the models in both steps have same network structure and we test both Convolutional Neural Networks (CNN) and Recurrent Neural Networks (RNN). Our pre-processing method uses light curves and period of stars as input data. We consider most of the classes and subclasses of variable stars in OGLE-IV database and show that using Hierarchical Classification technique and designing appropriate pre-processing can increase accuracy of predicting smaller classes, ACep and T2Cep. We obtain an accuracy of $98\%$ for class classification and $93\%$ for subclasses classification.
\end{abstract}

\begin{keywords}
  Variable Stars, Hierarchical Method, Convolutional Neural Networks
\end{keywords}

\section{Introduction}\label{sec:introduction}
Variable Stars are important objects in astrophysics and cosmology. Most importantly, there is well-defined relationships between the period and absolute magnitude of Cepheid stars, which are relatively young, massive and radially pulsating stars. These stars are used as the standard candles for measuring the cosmological distance and calibrating type Ia supernovas \cite{Freedman_2001,Riess_2011}.\\
\indent The Cepheids as the distance indicators also provide  essential information on the size of our galaxy \cite{Catelan_Smith_2015}. Type II Cepheids are also useful for studying stellar evolution as each subclass of this class is in a different stage of stellar evolution \cite{Soszynski_2010}. RR Lyrae stars can be used for studying the early history of galaxies. They are one of the oldest observable populations of stars and the chemical and dynamical evolution of this type of stars provides better understanding on the evolution of stars \cite{Soszynski_2009}.\\
\indent In order to classify variable stars, astronomers construct a set of features to describe light curves and extract them using Certain algorithms. Debosscher, J. et al. \cite{Debosscher_2007} used the Lomb-Scargle periodogram to find 28 properties of variable stars like median and mean of magnitude of light curves. Kim,
Dae-Won et al. \cite{Kim_2014} classified EROS-2 LMC variable stars by using Random Forest (RF) algorithm, using multiple features extracted from light curves. The features included Period, period S/N, Color and magnitude among others which were extracted using algorithms like Lomb \cite{Lomb_1976} and Scargle \cite{Scargle_1982}. They tested more than 30 features and selected 22 features to train their classifier based on the feature importance. They also computed the feature importance and found out that the period of variable stars is the most important property extracted in classifying by RF.\\
\indent Nun et al. \cite{Nun_2015} published the FATS package to extract features from photometry data automatically.  Kim \& Bailer-Jones \cite{Kim_2016} designed the UPSILON package, which extracts the features and then predicts the class of variable stars. This package works with a RF classifier which was trained with OGLE and EROS-2 data and its performance has been tested on MACHO, LINEAR, and ASAS database. All these works use sophisticated pre-processing algorithms to extract relevant information of the variable stars which leads to high computational cost. This is not practical for large cosmological datasets such as the Large Synoptic Survey Telescope (LSST) which will gather approximately 20 TB of data every night \cite{Ivezi_2019}. For realistic applications of these models on large astronomical data, it is essential to use efficient and low-cost pre-processing techniques.\\
\indent In the light of recent advances in Deep Learning, Neural Networks make a good candidate for classification of Variable Stars. Using neural networks can help saving time by using transformed data instead of extracted features for networks' input. Transformed data here refers to data which is reshaped to become appropriate as input, without any complex calculations like algorithms used for feature extraction.\\
\indent Aguirre et al. \cite{Aguirre_2019} and Becker et al. \cite{Becker_2020} proposed models for classification of OGLE-III variable stars using Convolutional Neural Networks (CNN) and Recurrent Neural Networks (RNN), respectively. They did not use any feature extraction. Instead, they transformed the light curves into matrix representations with differences in time and magnitude as elements. In the case of data, Aguirre et al. \cite{Aguirre_2019} did not consider some classes like T2Cep and ACep. Also, Becker et al. \cite{Becker_2020} accounted T2Cep class as Cep class and ignored Acep variable stars. These are classes with smaller population and they are usually challenging to classify.\\
\indent Machine learning models tend to train poorly when the training data is imbalanced i.e, distribution of stars in classes is biased. For such problems, machine learning models tend to over-fit to more populated classes and ignore smaller classes. For instance, in the case of variable stars, ACep makes only $~0.83\%$ of samples. It means that if the model ignores the ACep completely and classifies all the ACep as any other class, this would lead to an error of less than $~1\%$ based on accuracy. In many cases, we are interested in the less populated classes which correspond to rare events. For instance, as was explained before, detection/identification of ACep and T2Cep are of particular interest.\\
\indent In recent years, hierarchical methods have come to notice for classification by machine learning tools.  S{\'a}nchez-S{\'a}ez et al. \cite{Sanchez_2021} used hierarchical classification to classify several objects in the sky, especially variable stars. They selected 152 features for each star in classification, mostly defined from previous works like Kim, Dae-Won et al. \cite{Kim_2014} and
Nun et al. \cite{Nun_2015}. Rimoldini et al. \cite{Rimoldini_2019} also used hierarchical classification with 5 steps which uses different features in each step to distinguish stars.\\
\indent We use hierarchical technique and train our Neural Networks using OGLE-IV database \cite{Udalski_2015}. We only use the I-band photometry of the stars, i.e., time series containing the magnitude at different observation times and the period of each light curve \cite{Graham_2013}. For comparison, we have also trained an RNN based model. We find that the CNN has a lower computational cost and better performance, especially for less populated classes.\\
\indent Knowing what subclass a star belongs to, can be of importance. Therefore, the structure of our model consists of two general steps of predicting class and then subclass of the stars.\\

\indent They provide information about stellar evolution. For instance, each subclass of Type II Cepheids is in a very short state between two steps in stellar evolution, located in the separated areas in the Hertzsprung-Russel diagram. Because of their small population, some of these subclasses cannot be classified without using Fourier transform or other period related information about them \cite{Rimoldini_2019}. So, in this work, we use the period of stars to obtain phase-folded light curves as input in addition to using the period itself.\\
\indent This paper is organized as follows. Section 2 presents the dataset used for training and testing. In section 3, we propose the method designed for pre-processing. Section 4 presents the hierarchical classification and the classifier used in this technique. In section 5, we compare our results to some of the papers on this subject, and we also specifically consider results of less populated classes. Finally, in section 6, we present the conclusion and future work.
\begin{table*}
\begin{center}
\caption{This table includes information on each class of the data used in this work. Frequency is the number of is the ratio of the number of stars in each class to the total population of stars.}
\label{tab:Classes}
\begin{tabular}{lccr}
\hline
        Class (Acronym) & Num. Of Subclasses  & Num. of stars  & Frequency(\%) \\ 
\hline
        Eclipsing and Binary Stars (ECL)   & 2 & 9945 & 22.76 \\
        RR Lyrae (RRLYR)                   & 2 & 9519 & 21.79 \\
        Long Period Variables (LPV)        & 3 & 9914 & 22.69 \\
        Delta Scuti (DSCT)                 & 1 & 2678 &  6.13 \\
        Cepheid (Cep)                      & 3 & 9478 & 21.69 \\
        Type II Cepheid (T2Cep)            & 3 & 1783 &  4.08 \\
        Anomalous Cepheid (ACep)           & 2 &  366 &  0.83 \\
\hline
		Total & 16 & 43683 & \\
\hline
\end{tabular}
\end{center}
\end{table*}
\begin{table*}
\begin{center}
\caption{This table contains the number of each subclass in the OGLE dataset used for training, validation, and testing the performance of neural networks. Frequency is the ratio of the number of stars in each subclass to the total number of variable stars.}
\label{tab:SubClasses}
\begin{tabular}{llcr}

\hline
        Class & Subclass & Num. of stars & Frequency(\%) \\
\hline
        ECL   & NC     & 8209 & 20.30 \\
	    ECL   & C      & 1736 &  4.29 \\
	    RRLYR & RRab   & 7095 & 17.54 \\
	    RRLYR & RRc    & 2424 &  5.99 \\
	    LPV   & Mira   & 188  &  0.46 \\
        LPV   & OSARG  & 8455 & 20.91 \\
        LPV   & SRV    & 1271 &  3.14 \\
        DSCT  & SINGLE & 2678 &  6.62 \\
        Cep   & F      & 5315 & 13.14 \\
		Cep   & 1	   & 3469 &  8.58 \\
		Cep   & 12     & 694  &  1.72 \\
        T2Cep &	BLHer  & 747  &  1.85 \\
		T2Cep & RVTau  & 346  &  0.86 \\
		T2Cep & Wvir   & 690  &  1.71 \\
		ACep  & F      & 246  &  0.61 \\
        ACep  & 1      & 120  &  0.30 \\
\hline
        Total &        & 43683 &      \\
\hline
\end{tabular}
\end{center}
\end{table*}
\section{Data}\label{sec:Data}
We used the OGLE-IV \cite{Udalski_2015} variable stars database for training and testing \footnote{Other references related to the OGLE database : \cite{Udalski_2008,Soszynski_2014,Soszynski_2015,Soszynski_2015_2,Soszynski_2016,Soszynski_2016_2,Pawlak_2016,Soszynski_2017,Soszynski_2017_2,Udalski_2018,Soszynski_2018,Soszynski_2019,Soszynski_2020}}. Data contains 7 classes and 16 subclasses in total (see Table~\ref{tab:Classes} and Table~\ref{tab:SubClasses} for the information of the classes and sub-classes used in this paper).\\
\indent For each star, the data includes observation time in Julian days, magnitude and error bar of magnitude. The initial data has fluctuations due to the photometric error, the periodical nature of variable stars and some gaps due to sampling of the light curves. In this database, the population of different classes are not balanced. Machine learning techniques tend to favor more populated classes. This means that probability of misclassification increases for stars of less populated classes. To solve this problem we implement two solutions. The first is balancing the data, and the second is using the hierarchical technique presented in section \ref{sec:models}. To balance the population of classes, a maximum of 10,000 random stars was chosen from each class. Also, we removed some subclasses with less than 15 stars, like pWVir and BLHer1O subclasses from T2Cep. We used a total of 43,683 stars from the OGLE database to train and test the neural network model. $70\%$ of the data were used for training, $10\%$ for validation, and $20\%$ for testing the performance.
\begin{figure*}
    \centering
    \includegraphics[width=\columnwidth]{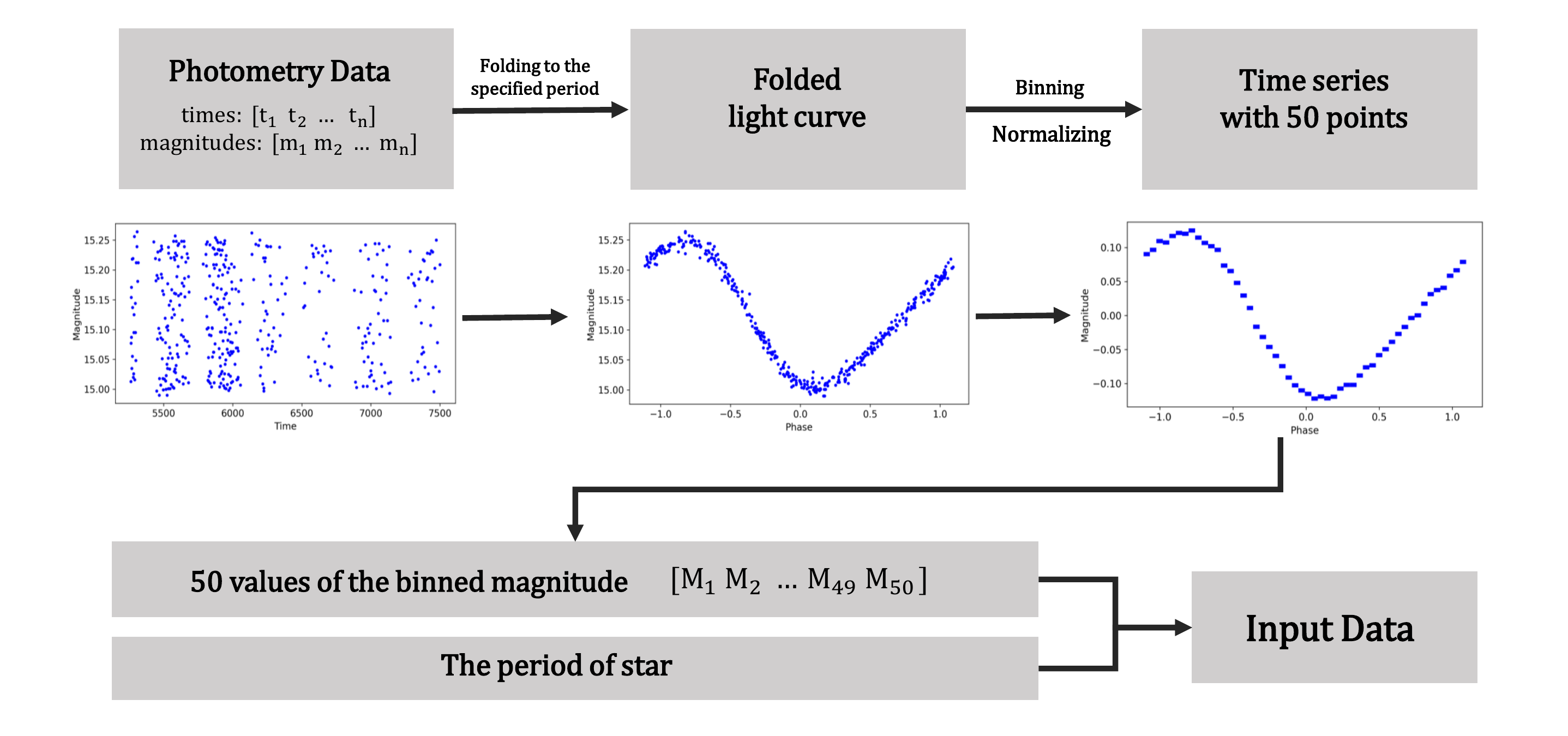}
    \caption{Schematic of pre-processing steps. In the first step, we take the raw data from the OGLE-IV dataset. Then in the second step, we fold raw data by using its specific period provided by the OGLE catalogue. We try binning the data to make the data with same length in the third step. Then, we add the period as an important feature to the input data. For an example, pre-processing steps on OGLE-LMC-CEP-0004 is shown. For details of pre-processing, see section \ref{sec:Pre-processing}.}
    \label{fig:Figure_1}
\end{figure*}
\section{Pre-processing input data}\label{sec:Pre-processing}
The raw data from different surveys differ in the sampling rate and even in the number of observation points due to complications such as the weather and moon's effects. To make the models independent of sampling rate, the pre-processing method produces normalized and binned light curves.\\
\indent To start pre-processing, first we use the period of variable stars, which is provided by the OGLE catalogue to fold the raw data using the ``lightkurve'' package \cite{Astropy_2018}. This step leads to obtaining the periodic behavior of the star in the phase space. Then, each phase folded light curve is divided into 50 equal bins to make the length of data points the same, and the value of each bin is set to the average of the values of the points in it. During this process, some bins become empty because there are no data points in them. To address this problem, we replace the empty values using linear interpolation.\\
\indent To find the best number of bins, we should consider that increasing the number of bins leads to increasing empty values, and decreasing the number of bins results in ignoring the details of the light curve. We test 25, 50, and 75 for the number of bins, and we find that 50 gives the best classification result. We also normalize the light curve by the mean of its magnitude. This would enhance the performance of the neural network model.\\
\indent The folded light curve has the disadvantage of losing period information which is a critical feature for classification of variable stars. To solve this problem, the value of the period is provided as a separate feature to the model. Fig.\ref{fig:Figure_1} shows the pre-processing pipeline.
\begin{figure}
    \centering
    \includegraphics[scale=0.33]{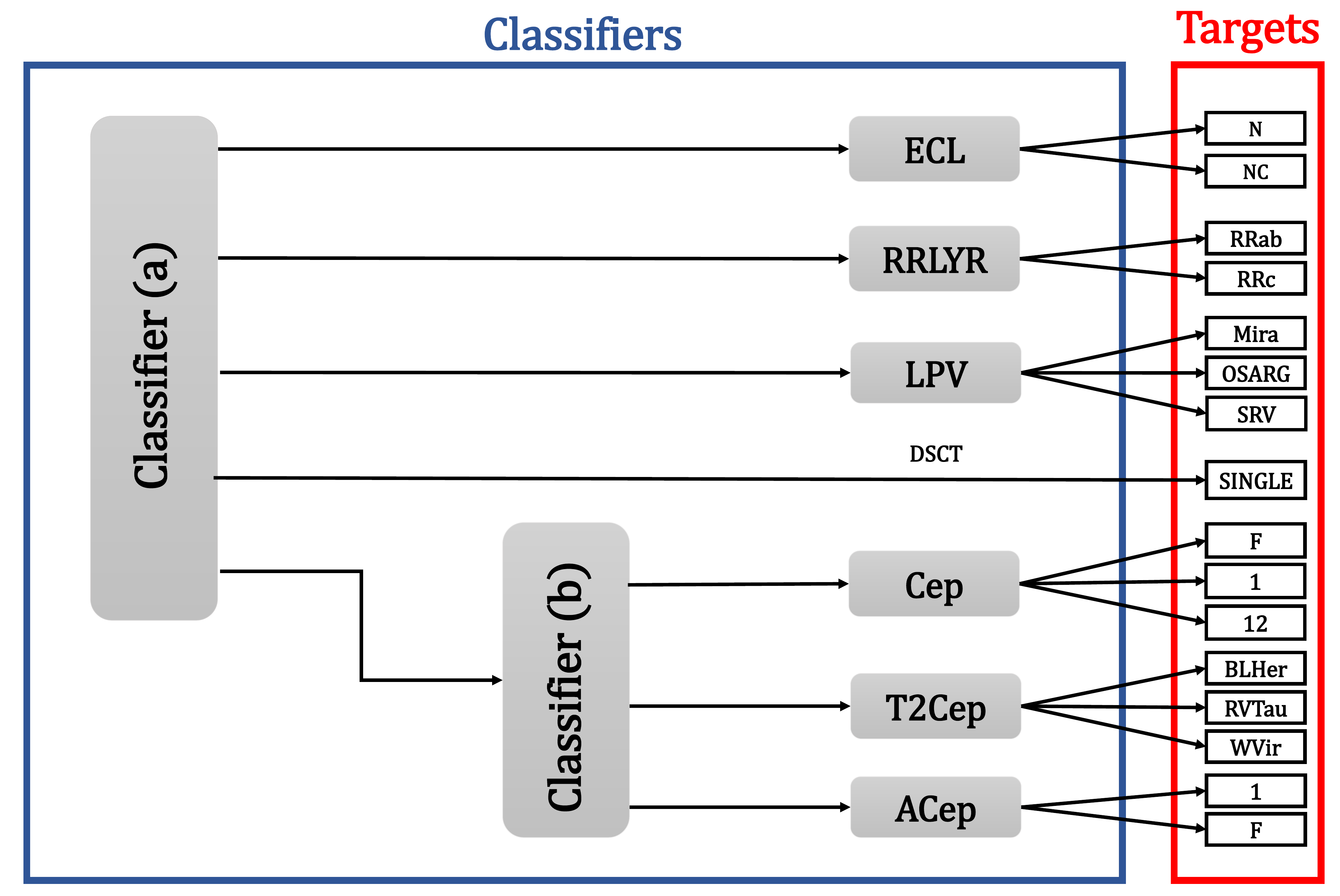}
    \caption{Schematic of Hierarchical Classification. Hierarchical classifiers are shown in the blue box. Classifier(a) and Classifier(b) are built to predict class of the stars, and the other classifiers named by their classes are designed to predict subclasses. The red box shows the targets.}
    \label{fig:Figure_2}
\end{figure}
\section{Model}\label{sec:models}
\begin{figure}
    \centering
    \includegraphics[scale=0.4]{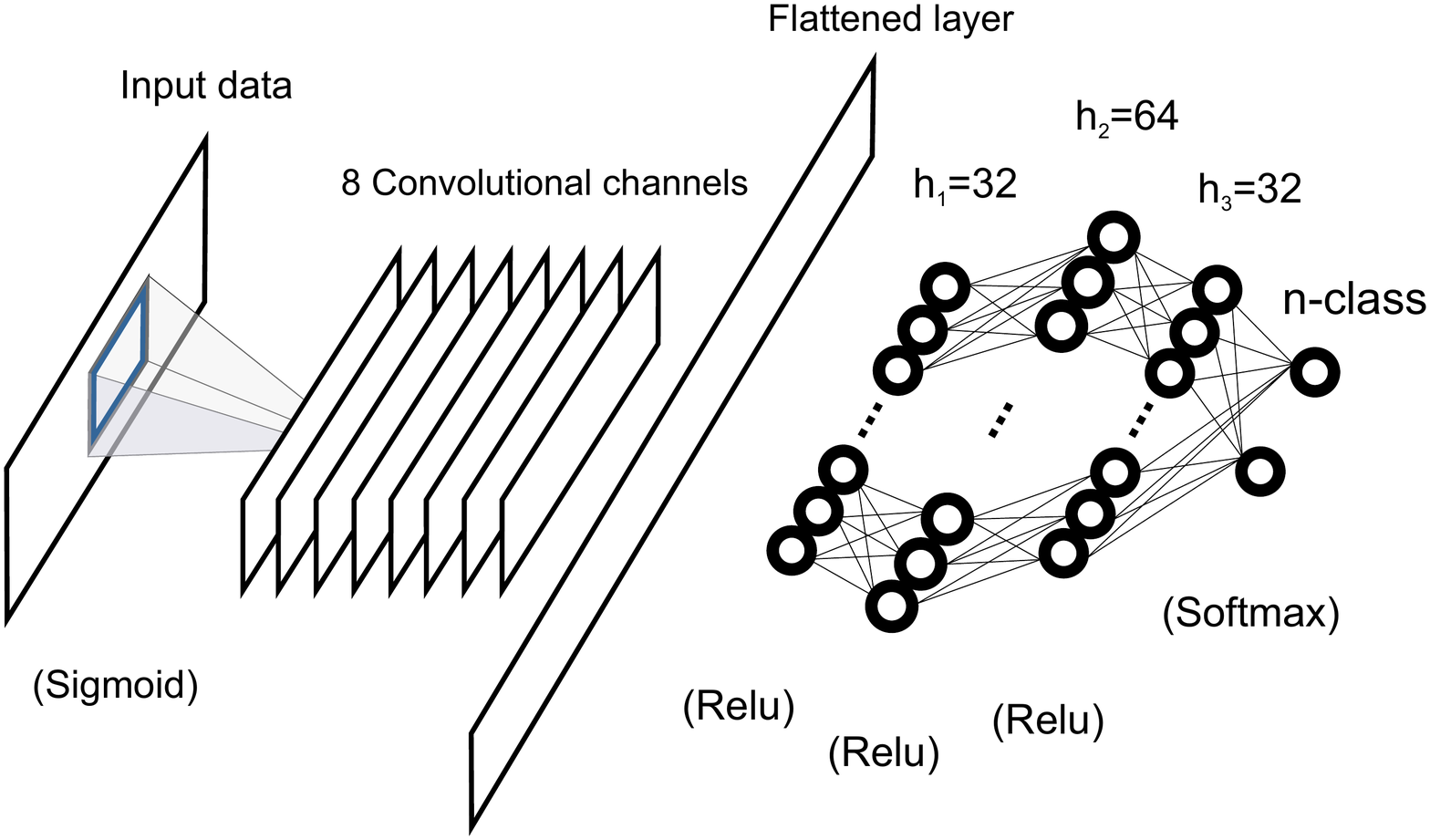}
    \caption{Schematic of Convolutional Neural Networks. In a simple neural network, $h_i$ is the number of nodes in each layer. In the final layer, n-class is the number of classes which the model should classify, e.g., 2 for the ECL classifier. The words in parenthesis are the name of activation functions, e.g., Relu. Blue rectangular is a filter that makes a convolutional layer.}
    \label{fig:Figure_3}
\end{figure}

We use a Hierarchical classification model technique, i.e., we use several classifiers to achieve better accuracy. The first classifier which we refer to as Classifier(a) separates ECL, RRLYR, LPV, DSCT, and a 5th class containing Cep, ACep, and T2Cep classes. More specifically, for the first level of classification, we group the three classes that are less populated and similar together. When the first classifier outputs the 5th class, a second classifier is used to specify which subgroup, i.e.,  Cep, ACep or T2Cep, the star belongs to. Up to this point, the combination of the two classifiers makes a classifier that can identify the class of the variable stars. Next, for each class that is composed of subclasses, a new classifier is trained to identify them.\\
\indent A schematic picture of the structure of our hierarchical classification model is depicted in Fig.~\ref{fig:Figure_2} (acronyms used for classes are introduced in Table \ref{tab:Classes}). This is in contrast to models used in \cite{Becker_2020} where one classifier is directly trained to classify subclasses, and then they group the subclasses to find the classification result for classes. Having several classes with different subclasses increases the probability of misclassification. The Hierarchical Classification technique helps to reduce these errors when we deal with multi-class classification. Specifically, this structure limits instances where stars from less populated classes such as ACep are assigned to more populated classes like RRLYR.\\
\indent The classifiers used for the hierarchical model are identical in the network structure. The model is composed of a 1D convolutional layer with eight channels followed by three fully connected layer. We used the ``Keras'' library \cite{Chollet_2018} to design our classifiers.\\
 \indent CNN are a type of neural networks that are often  used for image processing. CNN layers exploit properties in the input data such as locality or transnational in-variance for parameter sharing \cite{CNN_2017}. This reduces the number of parameters of the neural network which reduces the training costs and also makes the model less susceptible to over-fitting \cite{CNN_2017}. The schematic structure of the network is shown in Fig.~\ref{fig:Figure_3}. As shown in the figure, the model makes 8 convolutional channels from input data. Then, the flattened layer make appropriate data for the fully connected neural network which have three layers with different nodes. For details of CNN model, see Table.~\ref{tab:CNN_Inf}.\\
\indent For comparison, we also try another type of classifier, which is known as RNN \cite{RNN_2015}. The RNN models are typically used for classification of sequential data like the photometry data. This is why one may expect to get better performance from RNN. In this work, we use RNN models with two layers of simple RNN with 3 and 32 units, respectively. Also, we add two dense layers to increase the number of trainable parameters to have flexible models. See Table.~\ref{tab:RNN_Inf} for more information. However, we find that RNN does not offer any advantages over the CNN-based models and in fact, CNN model provides a faster classification with the same accuracy.
\begin{table}
    \centering
	\caption{Comparing accuracy and runtime of each type of neural network used to classify variable stars.}
	\label{tab:runtime_performance}
    \begin{tabular}{c c c }
        \hline
        Model & Accuracy & Training Time  \\
              &          & (min)           \\
        \hline
        CNN & 0.93    & $\sim$ 24     \\
        RNN & 0.90    & $\sim$ 324    \\
        \hline
    \end{tabular}
\end{table}
\begin{table}
	\centering
	\caption{Classification results of less populated subclasses using CNN and RNN classifier in hierarchical model.}
	\label{tab:results_CEPs}
	\begin{tabular}{l l c c} 
		\hline
		Class & Subclass & CNN & RNN \\
		    &   & Precision-Recall & Precision-Recall \\
		\hline
        Cep   & F     & 0.98 - 0.98 & 0.97 - 0.90 \\
        Cep   & 1     & 0.87 - 0.98 & 0.82 - 0.90 \\
        Cep   & 12    & 0.72 - 0.45 & 0.67 - 0.28 \\
        T2Cep & BLHer & 0.94 - 0.88 & 0.90 - 0.83 \\
        T2Cep & RVTau & 0.86 - 0.87 & 0.84 - 0.81 \\
        T2Cep & Wvir  & 0.95 - 0.89 & 0.75 - 0.95 \\
        ACep  & F     & 0.92 - 0.98 & 0.48 - 0.98 \\
        ACep  & 1     & 0.43 - 0.25 & 0.29 - 0.67 \\
		\hline
	\end{tabular}
\end{table}
\section{Classification results}\label{sec:result}
\indent We trained and tested our  models on Google collaboratory, which is an online programming environment equipped with 12 GB of RAM and NVIDIA Tesla T4 GPU.\\
\indent We have tested different models (CNN, RNN). For evaluation, we consider two main metrics, the accuracy of classification and the time required for training models. Table \ref{tab:runtime_performance} shows the accuracy and time cost of our models. This shows that the CNN model has better performance and is faster.\\
\indent One of the key aspects of our model is its ability to classify classes and subclasses with smaller populations. To evaluate the performance for unbalanced datasets like ours, accuracy is not enough. We also need to measure how well samples from smaller classes are detected. For this, we use ``recall'' which for each class, is the fraction of the samples from that class that have been correctly classified. For instance, the Cep has $1916$ samples in the test set and our model classifies $1863$ of them as Cep which means that its recall for Cep is $0.97$.\\
\indent Another metric that should be considered for imbalanced datasets, is ``precision'' which is the fraction of samples that are correctly predicted for each class. For instance, our model predicted $1907$ samples from the test set to be Cep, out of which, only $1863$ stars are actually Cep. This means that the precision for the Cep class is $0.97$. 
The performance metrics are discussed in more details in appendix \ref{sec:Performance Metric}.\\
\begin{table}
	\centering
	\caption{Results of classes classification using CNN models.}
	\label{tab:results_Classes}
	\begin{tabular}{l c r} 
		\hline
		Class &  Precision & Recall\\
		\hline
	    ECL   & 1.00 & 0.98 \\
        RRLYR & 0.98 & 0.99 \\
        LPV   & 0.98 & 1.00 \\
        DSCT  & 0.99 & 1.00 \\
        Cep   & 0.98 & 0.97 \\
        T2Cep & 0.94 & 0.90 \\
        ACep  & 0.82 & 0.74 \\
     \hline
       weighted avg & 0.98 & 0.98  \\
		\hline
	\end{tabular}
\end{table}
\begin{table}
	\centering
	\caption{Results of subclasses classification using CNN models.}
	\label{tab:results_SubClasses}
	\begin{tabular}{l l c r} 
		\hline
		Class & Subclass &  Precision & Recall \\
		\hline
        ECL   & NC     & 0.91 & 0.92 \\
        ECL   & C      & 0.68 & 0.57 \\  
        RRLYR & RRab   & 0.98 & 1.00 \\
        RRLYR & RRc    & 0.96 & 0.98 \\
        LPV   & MIRA   & 1.00 & 0.86 \\
        LPV   & OSARG  & 0.96 & 0.97 \\
        LPV   & SRV    & 0.79 & 0.84 \\
        DSCT  & SINGLE & 0.99 & 1.00 \\
        Cep   & F      & 0.98 & 0.98 \\
        Cep   & 1      & 0.87 & 0.93 \\
        Cep   & 12     & 0.72 & 0.45 \\
        T2Cep & BLHer  & 0.94 & 0.88 \\
        T2Cep & RVTau  & 0.86 & 0.87 \\
        T2Cep & Wvir   & 0.95 & 0.89 \\
        ACep  & F      & 0.92 & 0.98 \\
        ACep  & 1      & 0.43 & 0.25 \\
     \hline
        weighted avg & & 0.93 & 0.93 \\
		\hline
	\end{tabular}
\end{table}
\indent Table \ref{tab:results_CEPs} shows the comparison of classification results for CNN and RNN based models for less populated classes. For the performance of all classes and subclasses for the CNN model see Tables \ref{tab:results_Classes} and \ref{tab:results_SubClasses}.\\
\indent These two tables show that the models have better results for more populated classes than smaller ones, based on precision and recall.\\
\begin{figure*}
\begin{center}
    \includegraphics[scale=0.5]{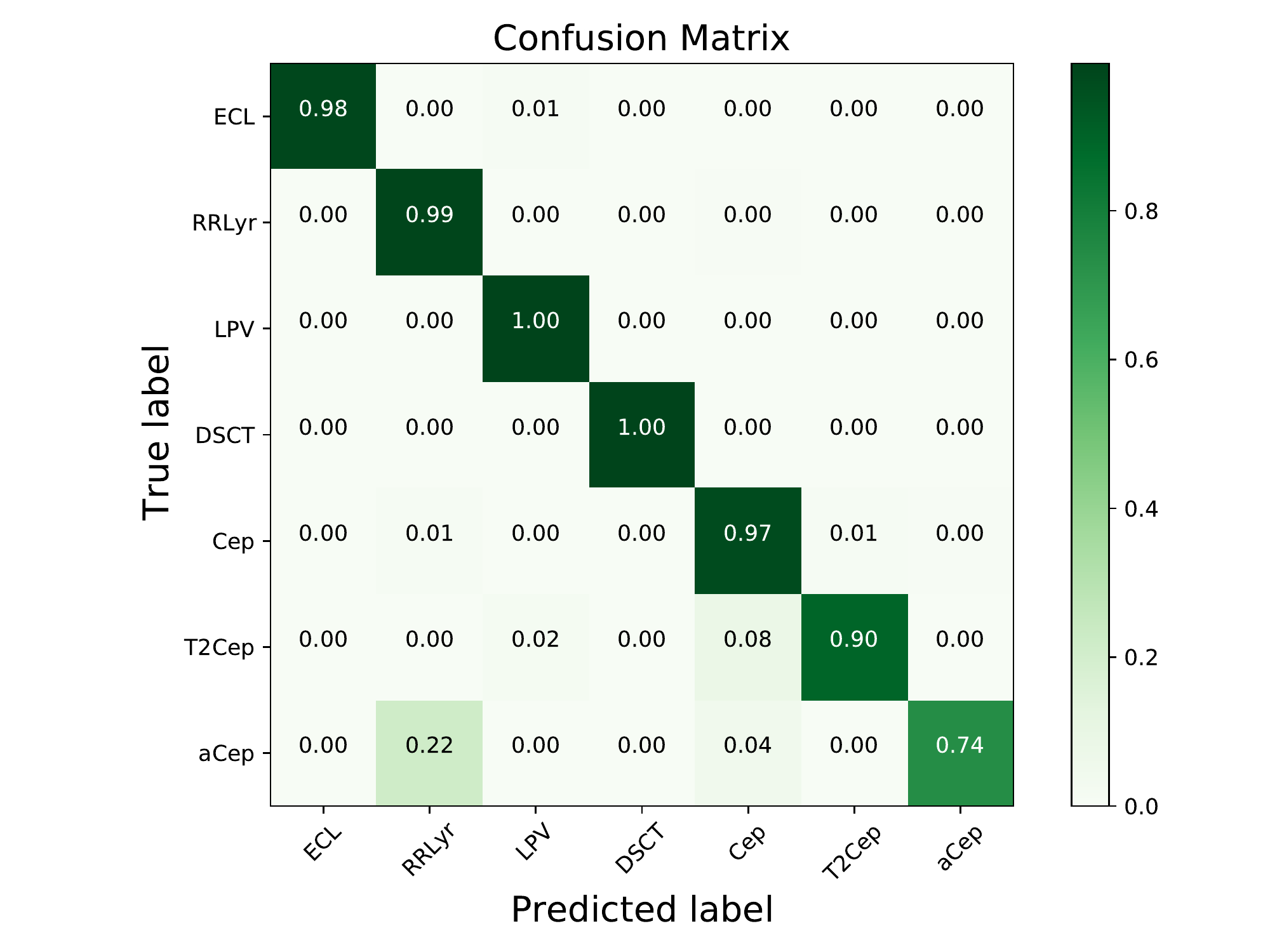}
    \caption{Confusion matrix of classification using CNN models and the pre-processing method discussed in section \ref{sec:Pre-processing}, only containing classes.}
    \label{fig:Figure_4}
\end{center}
\end{figure*}
\begin{figure*}
    \centering
    \includegraphics[width=\columnwidth]{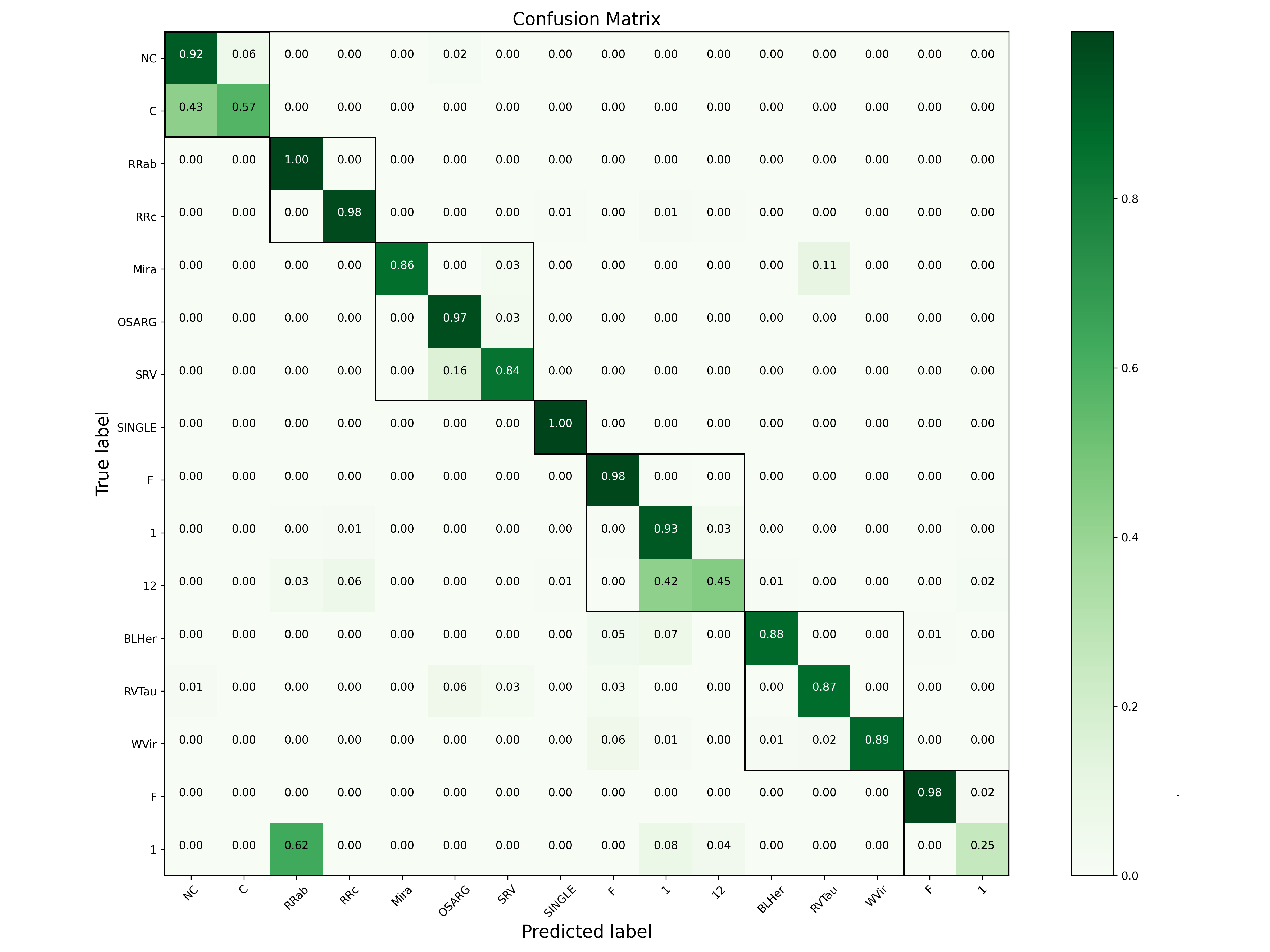}
    \caption{Confusion matrix of classification using CNN models and the pre-processing method discussed in section \ref{sec:Pre-processing}. The black squares highlight the seven classes of the top level (from top to bottom, ECL, RRLYR, LPV, DSCT, Cep, T2Cep and ACep, respectively)}
    \label{fig:Figure_5}
\end{figure*}
\indent In addition to mentioned metrics, we can evaluate our model by using confusion matrix. The Confusion matrix of classes and subclasses is shown in Fig. \ref{fig:Figure_4} and \ref{fig:Figure_5} for the CNN based model discussed in the paper.\\
\indent The more the matrix is diagonal, the more the classification is accurate. Checking the confusion matrix allows us to find the misclassifications made by the model. Therefore, the numbers out of the diagonal show misclassified classes and subclasses.\\
\indent Considering the confusion matrix, we find that misclassification in some subclasses are more than others. The main reason for these mistakes is the similarity of the light curves and the periods of the two subclasses. Subclasses with small population can be hard to predict, too, due to the data not being enough for training the network.\\
\indent As an example, $62\%$ of ACep (subclass 1) stars have been predicted as RRab stars. The reason for this is the similarity between period and light curve of these two types of stars. To be able to enhance prediction of ACep stars, information of distance and absolute magnitude is needed \cite{Soszynski_2008}.\\
\indent By comparing confusion matrices, we can find misclassifications in the second step of hierarchical classification. For instance, in Fig. \ref{fig:Figure_4}, $98\%$ of ECLs predict as actual labels; however, in the subclass confusion matrix (Fig. \ref{fig:Figure_5}), we find misclassifications in both subclasses of ECL, especially in the second one. This means class classifiers work well to distinguish  ECLs from other classes; however, subclass classifiers in the second step could not classify subclasses as good as the first step. This happens more in classes and subclasses with imbalanced data and leads to less accuracy for subclass classification than class classification.
\section{Conclusion}
We present a hierarchical model for classification of variable stars. The model has two neural networks for classification of the class of the star. Then each class is passed to a new neural network to identify the subclasses of the corresponding class. With this architecture we manage to classify less populated classes/sub-classes with high performance (See tables \ref{tab:results_Classes} and \ref{tab:results_SubClasses}).\\
\indent We use methods mentioned in section \ref{sec:Pre-processing} to make phase folded light curves suitable for input data. Although we obtain satisfying results for classes and subclasses classification, this pre-processing method is very challenging and high cost because of the processing needed for calculating the period. But it has a significant advantage. When using other methods based on features extracted from light curves, we usually face over-fitting when there are many features, and we should select important ones to solve this problem. But our models classify variable stars by the shape of light curves and their essential feature, the period.\\
\indent This work was done using OGLE-IV dataset. Next step could be including other datasets like WISE and Gaia.

\appendix
\section{Model Information CNN and RNN Models}
Details of CNN and RNN models are provided in Table. \ref{tab:CNN_Inf} and Table. \ref{tab:RNN_Inf}.
\begin{sidewaystable}
	\centering
	\caption{Details of structure of the networks used in CNN model.}
	\label{tab:CNN_Inf}
	\begin{tabular}{l c c c c} 
		\hline
		Type of layer & Output shape & Number of parameters & Activation function & More   \\
		\hline
        1D Convolutional & 48 $\times$ 8 & 56     & Sigmoid & Filters = 8 , Kernels = 3 \\
        Flatten          & 384         & 0      & -  & \\
        Dense            & 32          & 12320  & Relu & \\
        Dense            & 64          & 2112   & Relu & \\
        Dense            & 32          & 2080   & Relu & \\
        Dense            & n-class  & n-class $\times$ 32 + n-class    & Softmax  \\
		\hline
		\multicolumn{4}{l}{Total trainable parameters = 16568 + ( n-class $\times$ 32 + n-class )} &\\
		\hline
		\hline
		\multicolumn{2}{l}{Loss function = categorical-crossentropy} & \multicolumn{1}{c}{Optimizer = ADAM} & \multicolumn{2}{l}{Metric = categorical-accuracy}\\
		\hline
	\end{tabular}
	\hspace{3cm}
	\caption{Details of structure of the networks used in RNN model.}
	\label{tab:RNN_Inf}
	\begin{tabular}{l c c c c} 
		\hline
		Type of layer & Output shape & Number of parameters & Activation function & More   \\
		\hline
        Simple RNN & 50 $\times$ 3 & 18     & Sigmoid & Unit = 3, Return sequences = True\\
        Simple RNN &      32       & 1152   & Tanh    & Unit = 32  \\
        Dense            & 32          & 1056  & Relu & \\
        Dense            & n-class  & n-class $\times$ 32 + n-class    & Softmax  \\
		\hline
		\multicolumn{4}{l}{Total trainable parameters = 2226 + ( n-class $\times$ 32 + n-class )} &\\
		\hline
		\hline
		\multicolumn{2}{l}{Loss function = categorical-crossentropy} & \multicolumn{1}{c}{Optimizer = ADAM} & \multicolumn{2}{l}{Metric = categorical-accuracy}\\
		\hline
	\end{tabular}
\end{sidewaystable}
\section{Performance Metric} \label{sec:Performance Metric}
\begin{figure}
    \centering
    \includegraphics[scale=0.65]{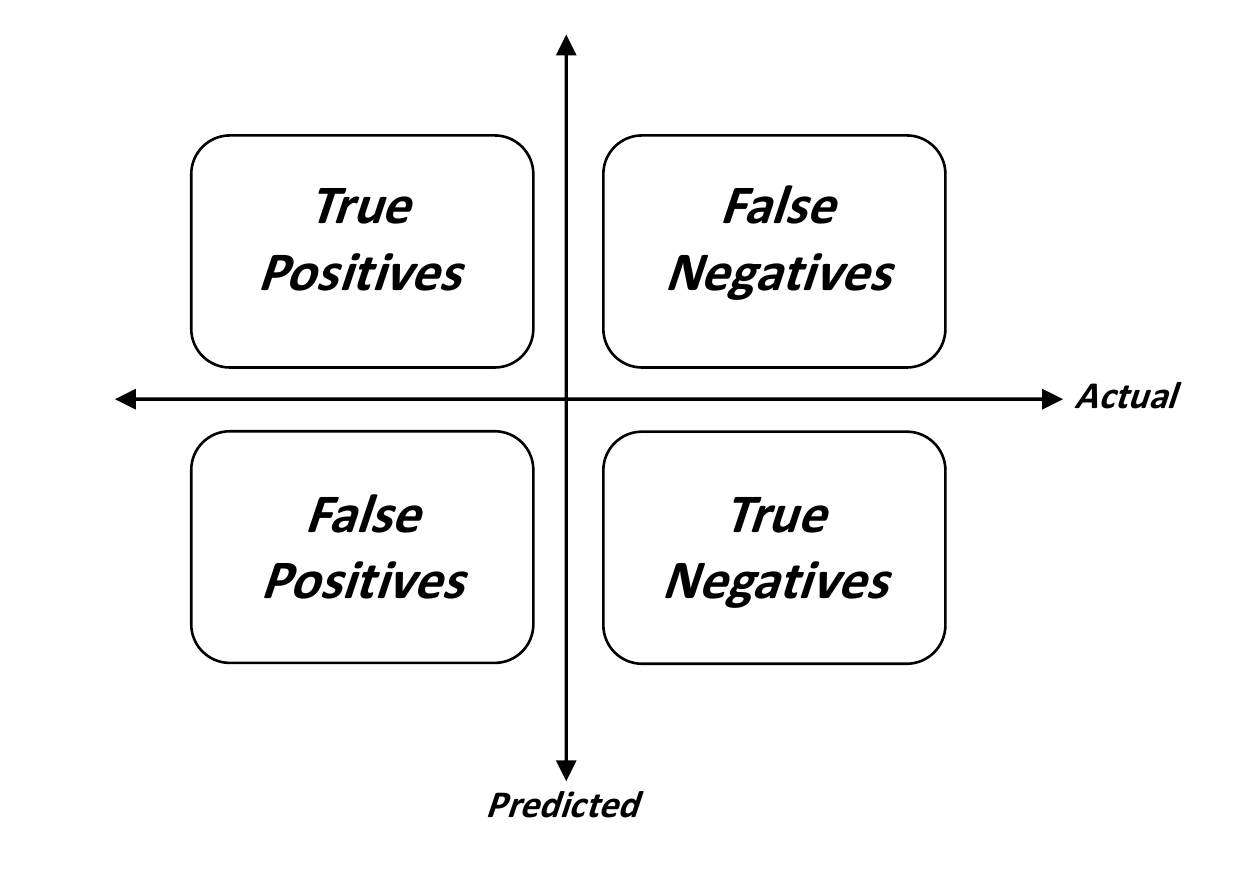}
    \caption{Illustration of true/false positives/negatives.}
    \label{fig:Apendix_Fig_1}
\end{figure}

The goal is to predict all classes and their corresponding subclasses as perfectly as possible. Because of the unequal population of the classes, accuracy could be misleading at metering the performance especially at the training step. We used precision and recall defined as follows (see Fig. \ref{fig:Apendix_Fig_1} for illustration):
\begin{equation} \label{eq1}
\textbf{\textmd{Precision}} =\frac{\textbf{\textmd{True Positive}}}{\textbf{\textmd{False Positive+True Positive}}},
\end{equation}
\begin{equation} \label{eq2}
\textbf{\textmd{Recall}}=\frac{\textbf{\textmd{True Positive}}}{\textbf{\textmd{False Negative+True Positive}}}.
\end{equation}
The weighted precision and recall are the precision and recall computed for each subclass separately and average weighted by the population of that subclass in the sample.

\end{document}